# Tracking Microstructure of Crystalline Materials: A Post-Processing Algorithm for Atomistic Simulations

Jason F. Panzarino[1], Timothy J. Rupert[1,2,3]

[1] Department of Mechanical and Aerospace Engineering, University of California at Irvine, Irvine, California 92697, USA

[2] Department of Chemical Engineering and Materials Science, University of California at Irvine, Irvine, California 92697, USA

[3] e-mail: trupert@uci.edu

**ABSTRACT**

Atomistic simulations have become a powerful tool in materials research due to the extremely fine spatial and temporal resolution provided by such techniques. In order to understand the fundamental principles which govern material behavior at the atomic scale and directly connect to experimental works, it is necessary to quantify the microstructure of materials simulated with atomistics. Specifically, quantitative tools for identifying crystallites, their crystallographic orientation, and overall sample texture do not currently exist. Here, we develop a post-processing algorithm capable of characterizing such features, while also documenting their evolution during a simulation. In addition, the data is presented in a way that parallels the visualization methods used in traditional experimental techniques. The utility of this algorithm is illustrated by analyzing several types of simulation cells which are commonly found in the atomistic modeling literature, but could also be applied to a variety of other atomistic studies which require precise identification and tracking of microstructure.

# INTRODUCTION

An important initiative within the materials science and engineering community has been integrated computational materials engineering (ICME), which aims to accelerate materials development and manufacturing processes by integrating computational materials science tools with experimental data and materials theory [1-3]. Since innovations in materials design and processing are often the driving force behind the development of advanced and disruptive technologies, rapid advanced material discoveries are essential for continued innovation. A key tenant of ICME is the creation of large databases of materials information, such as processing parameters, microstructure, and resultant properties, through joint computation and experimentation from which correlations can be drawn and new materials theory created. ICME relies on microstructure-mediated design, meaning quantitative characterization of three-dimensional (3-D) microstructures is of utmost importance [4].

The acquisition of large, 3-D datasets of microstructural features is challenging, although impressive progress has recently been made in the experimental community. A number of advances have come from uniting quantitative optical and scanning electron microscopy (SEM) with serial sectioning, allowing 3-D microstructures to be assembled by combining two-dimensional (2-D) scans from multiple oblique sections through a sample. Initial efforts in this field relied on manual material removal, with techniques such as milling or polishing [5], or focused ion beam (FIB) micromachining [6]. While manual sectioning is an excellent choice for fast material removal, FIB machining is very accurate and can produce sectioning with nanometer spacing. Echlin et al. have recently bridged the gap between manual and FIB-based sectioning by adding a femtosecond laser to a FIB/SEM system [7], creating a "TriBeam" system that can access a wide range of material removal rates [8]. In addition to serial sectioning techniques, electron tomography [9] and atom-probe tomography [10] have become increasingly popular for 3-D quantification of microstructure and both techniques can provide structural information with sub-nanometer resolution, but only limited volumes of material can be characterized. Unfortunately, there are inherent limitations to all of the experimental methods which are currently available. They are often very expensive, in terms of both equipment cost and the time needed to tabulate 3-D data sets, and usually destructive to the sample being analyzed. The first limitation hinders the accessibility to the broader materials community, while the latter means that one cannot track microstructural features as stress, temperature, or other driving forces for structural evolution are applied.

Atomistic simulations, such as molecular dynamics (MD) and Monte Carlo (MC) methods, can complement the available experimental techniques by providing a level of spatial and temporal resolution that experiments cannot achieve. Atomistic simulations also track atoms as the system evolves, making the documentation of system evolution a routine procedure. Even though current computational power forces MD timescales to be short and the spatial dimensions of MD and MC simulations to be small, there have been many examples where such methods have been successfully used to provide insight about material processing, microstructures, and properties [11]. For example, Wang et al. used MD to examine key morphological and compositional aspects of vapor-liquid-solid (VLS) growth of silicon nanowires [12]. Cheng and Ngan employed MD to study the sintering behavior of Cu nanoparticles, and found the process to be much different than what has been observed for larger particles [13]. Other examples of structural evolution documented with atomistic simulations include precipitate formation [14], film deposition [15], mechanically-driven grain growth [16, 17], phase transformations [18], and strain-induced amorphization [19], showing that atomistic modeling can be a powerful tool for documenting the processing-structure relations needed for ICME.

Unfortunately, the vast majority of characterization in atomistic modeling consists of the calculation of local properties of each atom, such as energy, stress, or local lattice distortion, and qualitative observations, such as the migration of a certain grain boundary. The quantification of microstructural evolution through rigorous feature tracking is less common. Luckily, a number of computational materials scientists have recently acknowledged this limitation and started working to develop tools for quantifying structure in atomistic simulations. For example, Stukowski and colleagues have created a dislocation extraction algorithm (DXA) that can identify both lattice [20] and grain boundary dislocations [21]. In addition, Xu and Li [22] have developed a technique for identifying atoms that are part of grain boundaries, triple junctions, and vertices, while Barrett et al. [23] and Tucker et al. [24] developed metrics which identified hexagonal basal plane vectors and microrotation vectors for all atoms, respectively. Some authors have gone an alternative route by simulating scattering physics in order to characterize microstructure, with Derlet et al. [25] and Coleman et al. [26] developing techniques for producing virtual diffraction profiles from atomistic data. In work that is closer to our end goal of analyzing crystalline materials directly, Tucker and Foiles recently developed a technique for finding individual grains within a polycrystalline sample, allowing for quantitative measurements of grain size [27].

Missing from the toolbox currently available to researchers is an analysis technique which can identify and track crystallites, their crystallographic orientation, and overall sample texture. In response to this need in the community, we have developed an original post-processing tool which identifies all crystalline grains and precisely calculates grain orientations with no a priori knowledge of the simulated microstructure. In addition, the algorithm also defines a mapping between simulation time steps, allowing for the analysis of individual grain movement, rotation, or coalescence as time progresses. In the present paper, we explain the details of our algorithm, while also providing several case studies showing its utility for characterizing and visualizing microstructural features in atomistic simulations.

## ANALYSIS METHODS

The *grain tracking algorithm* (GTA) presented in this paper consists of five principle steps:

(i) Crystalline atoms in the simulation set are identified by centrosymmetry parameter (CSP) [28], common neighbor analysis (CNA) [29, 30], or any other comparable measurement which can identify defects in local crystalline structure such as bond angle analysis [31] or neighbor distance analysis [32].

(ii) The local crystallographic orientation of each atom in a crystalline environment is calculated using the geometry of the material's unit cell.

(iii) Individual crystallites are identified by an iterative process where nearest neighbors must have similar crystallographic orientations to be included in the same grain.

(iv) Grains are indexed and tracked over time using the center of mass of each crystallite.

(v) The measured orientation of each grain and the overall sample texture are visualized with pole figures, inverse pole figures, and orientation maps. We aim to recreate the familiarity of experimental visualization methods and better integrate atomistic datasets into ICME.

It is important to note that the GTA is largely a generic algorithm that can be applied to any crystalline material regardless of the crystal structure. In this paper, we describe the algorithm's implementation for face centered cubic (fcc) crystals, although only minor modifications to Steps (ii) and (v) would be necessary to analyze other crystal structures. The algorithm is currently implemented as MATLAB code, which is available from the authors upon request.

*Atom Classification*

The first step in our GTA method is the separation of atoms based on their local environment. Specifically, we classify atoms as *grain interior*, *grain edge*, or *non-crystalline*. Many techniques for local structural analysis exist, and a detailed discussion of the advantages of each method can be found in a recent article from Stukowski [32]. While any of these metrics can be used with our algorithm, we focus here on CSP and CNA as potential techniques for defect identification, due to their widespread usage in the literature and because they are built into common MD simulation packages such as the Large-scale Atomic/Molecular Massively Parallel Simulator (LAMMPS) code [33].

A centrosymmetric lattice, such as fcc or body centered cubic (bcc), has pairs of equal and opposite bonds between nearest neighbors. CSP measures the deviation from this perfect centrosymmetry and can be used to identify defects when a threshold value associated with thermal vibrations is exceeded. One method for determining an appropriate threshold for defect identification is the Gilvarry relation [34], which places an upper limit on the thermal vibrations a crystal can experience before it melts (~12% of the nearest neighbor distance), but such a distinction is not perfect and the CSP method struggles with false positives in defect identification at high temperatures. However, the CSP metric is well-suited for analyzing highly strained atomistic systems, as it is not sensitive to homogeneous elastic deformation. Alternatively, CNA analyzes the topology of the bonds within a cut-off distance around an atom and assigns a structural type (fcc, bcc, hexagonal close packed (hcp), or unknown structure are the distinctions which are commonly used) to the atom in question. Therefore, any atom with a structure different than that expected for the material can be classified as a defect. For example, when analyzing fcc Ni, all bcc, hcp, and unknown atoms would be considered defects. CNA tends to be less sensitive to thermal vibrations, but large elastic strains can pull the nearest neighbors outside of the cut-off distance for analysis. Hence, the CNA metric is most useful when dealing with materials at high temperatures but struggles with highly strained systems.

Using either CSP or CNA in its current formulation, the GTA first identifies defect atoms and labels them *non-crystalline*. We then further sort the remaining crystalline atoms by examining the nearest neighbors. If all of an atom's nearest neighbors (12 in the case of fcc) are also in a crystalline environment, then the atom is labeled *grain interior*. Alternatively, if one or more nearest neighbor is non-crystalline, then the atom in question is labeled *grain edge*. This distinction ends up being important for identifying grains and avoiding large errors in the calculated orientation of the crystallites. Fig. 1 shows a polycrystalline Al atomistic sample, with atoms separated into grain interior, grain edge, and non-crystalline.

*Local Crystallographic Orientation*

Once all atoms in grain interior environments have been found, we calculate the local orientation of each atom based on the unit cell of the material. The process of determining the local orientation at each atom is highlighted in Fig. 2(a) for an fcc material. The nearest neighbors of an atom must be found, and then an arbitrary vector is chosen in the direction of one of these neighbors. For an fcc lattice, there will be four other neighbors whose directional vectors will lie approximately 60° from the original arbitrary vector. These four new vectors will reside in two separate {100} planes of the unit cell, with each plane containing a pair of nearest neighbors and a <100> direction which must be perpendicular to its counterpart. The cross product of these two <100> directions then gives the third <001> direction. Finally, we find the inverse of these three vectors as well, giving all six <100> axes of the unit cell. The local orientation of the atom is now described fully and can be stored. This calculation is repeated for each grain interior atom, providing crystal orientation as a function of position within the atomistic sample. It is important to note that any periodic boundary conditions must be enforced before this calculation, in order to avoid errors in atoms near the boundaries of the simulation cell. Our code takes care of this requirement by adding virtual images of the simulation cell when necessary.

While the exact description provided above is unique to an fcc lattice, only small changes are required for other lattice structures. For example, if the equilibrium crystal structure is bcc, then a similar method for finding the local orientation can be employed and is outlined in figure 2(b). We start by calculating the vectors connecting our atom of interest to its 8 nearest neighbors. Then, we can select two of the nearest neighbor vectors and their inverse vectors which are 180° away, giving four vectors which lie in a {110} plane. Next, we separate the pair of vectors which are 109.47° apart from the other pair that lies 70.53° apart. Adding the vectors in each pair gives the blue and orange axes shown in Fig. 2(b), and the cross product of these vectors gives the green axis. Because our vector algebra has been done on a {110} plane up until this point, we only need to rotate this new orthogonal coordinate system by 45° about the orange axis vector to find three <100> directions and obtain the center atom's local crystallographic orientation.

*Grain Identification*

After orientations are calculated for every grain interior atom, the GTA begins searching for and identifying individual grains. To begin, a randomly selected grain interior atom is picked and added to the current grain of interest as the first atom. This atom will temporarily be labeled as the *reference atom*. The nearest neighbors are then reviewed one by one and must meet certain criteria before being added to the grain currently being indexed.

The GTA first checks to make sure that the atom is also a grain interior atom. It is common for grains to be artificially connected by one or two atoms that are in a crystalline environment. Therefore, the segregation of grain edge atoms from the grain interior atoms which are deeper within the crystallite ensures more accurate grain identification by closing some of these artificial connections. Next, the orientation of our reference atom is compared with the orientations of its nearest neighbors, using a user defined *orientation-cutoff angle* as our metric. In its current formulation, the GTA calculates the angles separating the <100> directions associated with the reference atom and the <100> directions associated with the nearest neighbor in question. If all of these angles are less than the orientation-cutoff angle, the nearest neighbor atom is added to the grain. We currently use an orientation-cutoff of only a few degrees, which will be discussed more extensively in the Applications and Examples section of the text. The chosen orientation-cutoff angle can be adjusted for a variety of reasons, with an obvious example being the decision whether to identify low-angle grain boundaries or not. A low-angle grain boundary composed of an array of dislocations would not appear as a continuous plane of non-crystalline atoms and would therefore not be identified as a grain boundary if only CSP or CNA is used. However, the GTA recognizes the change in crystal orientation across such a boundary if the orientation-cutoff is low and allows the two grains to be distinguished from one another. Since the GTA stores all orientation information needed to completely describe each atom's local crystallographic environment, one could also choose to calculate alternative metrics, such as misorientation angle, to compare with the orientation-cutoff angle.

After examining the nearest neighbors of the first reference atom, the GTA then selects one of the atoms which was just added to the grain as the new reference atom and repeats the procedure for this atom's nearest neighbors. By repeating this process, the algorithm builds the current grain outward, identifying suitable atoms along the way, until it has found all of the atoms associated with the current grain. The GTA then calculates the center of mass and the average crystallographic orientation of the current grain, saving this data with the grain number. With the first grain complete, the GTA then selects another random grain interior atom, making sure that it has not already been checked and added to a grain, as the first reference atom for the second grain. After all grains are identified, a nearest neighbor search of the grain edge atoms is used to find which grain these atoms should be added to. This final step is important if one is interested in metrics such as grain size. While the grain edge atoms can be problematic for calculating orientation information, they are still crystalline and can be a significant fraction of the grain volume for very fine grained samples.

*Grain Tracking*

The GTA algorithm can analyze multiple output files from atomistic simulations and thus provide data regarding microstructural evolution through time. After all grains have been identified in each output file, the GTA then begins to identify and reassign each grain number such that it corresponds with its counterpart in the following time step. In order to accomplish this mapping, the center of mass of each grain in the initial reference configuration is compared to the next time step and the closest center of mass in the new file is found. Once these two grains are matched, the grain number of the new file is updated to match the grain number from the reference configuration. This process is then repeated for the remaining grains, until all grains are matched. While such a tracking mechanism can fail if a grain has moved too far away between successive output files, this problem can often be solved by simply analyzing the microstructure and tracking the grains more frequently during the simulation.

*Visualization Techniques*

In order to help facilitate the integration of the GTA into the combined computational-experimental framework needed for ICME, several common visualization techniques are employed by the algorithm. First, pole figures are developed by stereographically projecting a family of crystal axes for each grain with respect to a specified viewing direction. In the examples shown in this paper, we project the <100> poles. In order to simplify interpretation of the data, inverse pole figures can also be generated. Because of crystallographic symmetry for the fcc materials we focus on here, visualization of the inverse pole figure can be abbreviated into a single stereographic triangle. To produce these figures, the GTA automatically imposes all crystallographic symmetry operations for each grain and projects all associated poles stereographically. Those points which lie within the stereographic triangle are then plotted and graphically represent the orientation for each grain. Both of these methods are used extensively in the experimental community for visualizing texture. Finally, 3D orientation maps are also created by plotting all atom positions and color coding each grain according to its projected inverse pole. Such a visualization technique replicates traditional output of orientation imaging microscopy (OIM) software. All of these capabilities provide a direct link for simplifying the comparison of experimental texture data with those results produced by atomistic modeling.

## APPLICATIONS AND EXAMPLES

To illustrate the utility of the GTA as well as highlight user-controlled features and practical concerns for the algorithm, several common examples of atomistic samples were analyzed. MD simulations were carried out with the open-source LAMMPS code [33] using an integration time step of 2 fs, and embedded atom method (EAM) potentials for Ni and Al developed by Mishin et al. [35] were used. CSP is used to identify non-crystalline atoms, with $CSP \geq 2.14$ Å$^2$ and $CSP \geq 2.83$ Å$^2$ characterizing defects in Ni and Al, respectively. Additional simulation details will be given when necessary. All atomistic visualization in this manuscript was performed with the open-source visualization tool OVITO [36].

### *Effect of Temperature on a Ni Σ5 (310) Symmetric Tilt Grain Boundary*

We begin our analysis of atomistic examples by investigating a very simple, known sample microstructure: the Σ5 (310) grain boundary in Ni. The bicrystal sample shown in Fig. 3 was created by tilting the crystals around the [100] crystallographic axis until there is a misorientation of 36.87° between the top and bottom half. Fig. 3(a) shows this misorientation by drawing the <100> directions from each grain. Periodic boundary conditions were applied in the X- and Y-directions, while free boundary conditions were implemented in Z. Bicrystal samples such as these have been used extensively to investigate behavior such as dislocation emission from grain boundaries [37] or grain boundary migration [38]. These samples were equilibrated at zero pressure and temperatures of 10 K, 300 K, and 600 K for 20 ps.

Each sample was analyzed with the GTA using an orientation-cutoff of 3°, with results shown in Figs. 3(c)-(e). In these images, atoms are colored according to grain number, with light blue signifying the first grain (G1) and green showing the second grain (G2). Atoms which are not part of either grain are shown in dark blue. It is instructive to first focus on the sample at 10 K shown in Fig. 3(c), where thermal vibrations are very small. In this case, dark blue atoms only appear at the bicrystal interface and at the free surfaces, meaning all atoms inside of the grains have been properly indexed. Fig. 3(b) shows a {100} pole figure centered on the tilt axis of the bicrystal, or the X-axis of the simulation coordinates. While one <100> direction of each crystal is centered on the pole figure, the other <100> directions show the expected tilt rotation. As temperature is increased in Figs. 3(d) and (e), a significant number of atoms cannot be indexed to either grain. It is important to note though, that the average orientation we measure is unchanged by this noise.

At first glance, one might think that this behavior is simply the result of CSP artificially identifying atoms as being in a defective local environment. However, Fig. 3(f) shows atoms in the sample at 600 K colored according to CSP. Only a select few atoms within the grains are incorrectly identified as defects (white in this image), so this cannot explain the large number of dark blue atoms in Fig. 3(e). Closer analysis shows that these atoms are not assigned to a grain because their local crystallographic orientation is different than their neighbors' due to thermal vibrations. To highlight that the GTA is not sensitive to the choice of CSP or CNA, Fig. 4(g) shows the atoms colored according to CNA. While CNA has less trouble finding non-crystalline atoms at elevated temperature, we obtain the exact same result shown in Fig. 4(e) when we repeat the GTA analysis, again because of local orientation fluctuations due to temperature.

Whether or not this issue needs to be addressed will depend on the information that is of interest for the particular application. For example, if finding the average orientation of each grain is the main goal, then the false negatives inside the grain can be ignored and a restrictive orientation-cutoff angle can still be used. However, if it is necessary to accurately track grain size over time, all of the atoms inside the grain must be counted. One potential solution would be to run a subtle energy minimization procedure on the computational sample before analyzing with the GTA. Such a minimization will remove the noise from thermal vibrations, but care must be taken to ensure that it is not aggressive enough to significantly change larger features of the microstructure being analyzed. For all studies in this manuscript, we deemed a minimization to be appropriate if no significant orientation changes to the grains occurred. Justification of our energy minimization tolerance is further discussed in the next section. Fig. 3(h) shows the 600 K sample which was minimized with the conjugate gradient method in LAMMPS (using a unitless energy tolerance of $10^{-6}$ and a force tolerance of $10^{-6}$ eV/Å) and then analyzed, showing that all atoms in the grains are identified. Alternatively, a user can increase the orientation-cutoff angle to a larger value. Fig. 3(i) shows the 600 K sample analyzed again, but with an orientation-cutoff angle of 10°. In this case, all of the noise in local orientation induced by thermal vibrations is less than the cutoff value and all of the atoms are correctly identified. It is worth noting that increasing the orientation-cutoff angle could artificially lead to the merging of two grains into one, a possibility that will be discussed further in the next section.

*Texture Analysis of Nanocrystalline Al Samples*

We envision that a major application of the GTA will be the analysis of texture in atomistic simulations. For example, texture could be tracked during simulations of film deposition or deformation in nanostructured materials. To show the power of the GTA for such analysis, we next analyze two common types of nanocrystalline samples which are commonly found in the literature. Nanocrystalline materials are promising structural materials due to their extremely high strength [39] and atomistic simulations are often used to study their deformation physics in either columnar grained [40, 41] or random polycrystalline samples [42, 43]. Columnar grained structures allow for easy viewing of dislocation-boundary interactions, while random polycrystalline samples are more realistic microstructures. A columnar grained sample was generated by creating 36 random grain centers on a hexagonal lattice, and then building crystallites with a common <100> axis and a random rotation angle around this axis for each grain. A random polycrystalline sample with 46 grains was created using a Voronoi tessellation construction modified to enforce a minimum separation distance between grain nucleation sites [44] and Euler angles that were randomly selected for each grain. Since simply filling space with atoms until grains impinge gives an artificial microstructure, conjugate gradient minimizations in LAMMPS (energy and force tolerances of $10^{-6}$) were applied to both samples to create fully dense simulation samples by letting the atoms relax slightly. Both samples have an average grain size of 5 nm and contain Al atoms, and periodic boundary conditions are applied in all directions.

We begin our discussion of these samples by using the GTA, with an orientation-cutoff angle of 1°, to analyze the columnar grained sample in more detail. Fig. 4(a) and (b) show the columnar grained sample in both its as-assembled state and after minimization, with atoms colored according to their grain number. It is clear that the as-assembled sample is not fully dense, as many grain boundaries contain small nanoscale voids, but minimization closes this porosity. A few grains coalesce, most notably the two at the top left and the three near the top right of the sample. These grains actually had very similar rotations around the <110> axis and were only artificially separated by porosity in the as-assembled sample. After minimization, the artificial boundary is removed and the crystallographic orientations are close enough that they are considered one grain.

Many unique grains are still identified after minimization, and their orientations are used to create the {100} pole figure shown in Fig. 4(c). The inner circle on the pole figure is from the (100) and (010) planes of each grain, which are 45° away from the X-axis, while the outer circle comes from the (001) planes, which are perpendicular to the X-axis. This same data is also presented in inverse pole figures for each of the simulation axes in Fig 5. It is clear that only {110} planes are pointing in the X-direction, and the zoomed image of the bottom right corner shows that minimization only leads to a very small deviation from the as-assembled condition where grains are exactly columnar. A maximum out-of-plane rotation of 0.1° is observed for the minimized sample, and most grains are altered much less than this. Figs. 5(b) and (c) show the inverse pole figures for the Y and Z simulation axes, and the orientations are restricted to the top borders of the stereographic triangle due to the columnar nature of the grains. These plots confirm that we can recreate the type of orientation datasets that a researcher would take away from experimental investigations.

To show the effects of different choices for the orientation-cutoff angle more clearly, we focus on the collection of three grains marked with a dashed circle in Fig. 4(b). These grains are shown in Fig. 6 for orientation-cutoff angles of 1°, 1.5°, and 2°, with atoms colored according to their grain number. With the original choice of a 1° cutoff angle, three distinct grains are found, even though the red and orange grains do not have a discrete plane of non-crystalline atoms between them. As the cutoff angle for analysis is increased to 2°, these two grains are now identified as one by the new measurement standard. We make no judgment about which is correct, since the decision to exclude low angle boundaries may be application dependent.

We next move our attention to GTA analysis of the random polycrystalline sample in Fig. 7. With no restrictions on the Euler angles that defined the orientation of each grain, we expect to have a random texture. Fig. 7(a) shows the random polycrystalline sample with the atoms colored according to grain number while Figs. 7(b) and (c) present a {100} pole figure and an inverse pole figure down the X-axis, respectively. Neither Fig. 7(a) nor (b) shows any discernible pattern, confirming the sample's random texture. As a final comparison between columnar grained and random polycrystalline atomistic samples, Fig. 8 shows orientation maps for the X-axis of the simulation cells. While the columnar grained sample (Fig. 8(a)) is entirely green and has only {110} planes pointing in the X-direction, the random polycrystalline sample (Fig. 8(b)) shows a mixture of colors and orientations. Note that the black region on the front face of the random polycrystalline sample is simply a region where a grain boundary is located at the cell boundary. It is clear that the GTA can provide quantitative measurements of sample texture, while also presenting the data in a way that is intuitive.

*Strained Polycrystalline Sample*

The real value of the GTA arises when the starting microstructure is unknown or when structural changes must be tracked over time. We finish our analysis of example problems by investigating grain rotation during the deformation of nanocrystalline Al at different testing temperatures. The random polycrystalline sample introduced above was first equilibrated at 600 K and zero pressure for 100 ps using a Nose-Hoover thermo/barostat. One sample was deformed at 600 K, while two others were cooled to 450 K and 300 K and then deformed. A constant cooling rate of 30 K/ps was used to lower the temperature of the system. Deformation was simulated by applying a uniaxial tensile strain ($\varepsilon$) along the Z-axis at a constant true strain rate of $5 \times 10^8$ s$^{-1}$ while keeping zero stress on the other axes using an NPT ensemble.

Fig. 9(a) presents the true stress-strain curves from the nanocrystalline Al samples tested at different temperatures. As temperature is increased, both yield strength and flow stress decrease. Such behavior has been reported previously (see, e.g., [45, 46]), but here we can quantify the temperature dependence of an important deformation mechanism: grain rotation. While a handful of previous reports have tracked the rotation of a few select grains during MD deformation of nanocrystalline metals, these always represented a fraction of the total grains in the sample [17, 46, 47]. Here, we track all grains in the sample and track changes to their orientation as a function of strain for different temperatures. We examine this structural evolution in 2% strain intervals up to 10% applied true strain. For the limited number of grains that rotate and coalesce to form larger grains or shrink and disappear, we track orientation for as long as possible. Fig. 9(b) presents the average rotation of grains towards the tensile axis to provide a measurement of rotation as a function of strain. Perhaps not surprisingly because of the increase diffusion at higher temperatures, there is significantly more grain rotation on average for the sample tested at 600 K than for the sample tested at 300 K. At 10% applied strain, grains in the 600 K sample have rotated ~50% more than grains in the 300 K sample.

Because we track individual orientations as well, we can focus on interesting grains. Fig. 10 presents inverse pole figures from the tensile axis for the three testing temperatures, with orientations shown at different strains. We only plot five grains here to simplify visualization. Some grains experience a slow but steady rotation, while others experience large changes in orientation within one 2% strain interval. In general, while the grains rotate more at elevated temperatures, each grain rotates in roughly the same direction at every temperature. For example, G5 moves up and to the left in all frames of Fig. 10. This suggests that the rotation direction is likely limited by the compatibility with surrounding grains and only the magnitude of rotation is strongly affected by temperature. In addition to a quantitative understanding of textural changes, the algorithm also keeps track of grain shape and center of mass. As such, individual grain tracking can be carried out in a much simpler fashion. Traditionally, in order to document the merging of two adjacent grains one might search manually through the atomistic sample for such a case and then track the movement and rotation using visual alignment of atomic planes. Since grain sizes and orientations are calculated automatically with the GTA, it is easy to identify which grains will merge and visualization of these deformation mechanisms can be conducted quickly. To illustrate, Fig. 11 shows magnified images of the tracking of 3 grains. The red grain slides into the page during the tensile test (disappearing from view), while the gold and blue grains rotate toward each other and coalesce when 6% true strain is applied.

## CONCLUSIONS

Atomistic modeling tools can potentially provide the enormous datasets of 3-D microstructural features that are essential for ICME efforts, but only if characterization of these simulations evolves from anecdotal observations to quantitative metrics. In this article, we have introduced a new post-processing algorithm that can be used to identify and track microstructural changes in crystalline materials during computational studies on the atomic scale. The GTA enables the quantitative characterization of grain size, grain orientation, and sample texture while also tracking these features as a function of time during simulations of dynamic behavior. This data is also presented in ways that are commonplace within the experimental community, such as pole figures, inverse pole figures, and orientation maps, in order to further connect computational and experimental research. In order to clearly illustrate the capabilities of the GTA, a number of common MD simulation cells were also analyzed. These examples show that:

- Atomistic orientation measurements on the atomic scale can be made by applying simple crystallographic analysis techniques to an atom's local environment. This local orientation can enable the identification of even notoriously difficult to extract low-angle grain boundaries. By taking average orientations from all atoms, the crystallographic texture of known test samples was confirmed, showing that the extremes of strong out-of-plane texture and completely random texture could be identified.

- The thermal vibrations in high temperature simulations may make it difficult to index certain crystalline atoms to the correct grain if a restrictive orientation-cutoff angle is used. While this does not affect the measured orientation in any meaningful way, it should be important for tracking grain size and can be addressed by a larger cutoff angle or energy minimization techniques. Care must be taken that any energy minimization damps out these vibrations but does not dramatically alter larger microstructural features.

- Grain rotation was measured in nanocrystalline Al as a function of applied strain for three different testing temperatures. Higher temperatures lead to more grain rotation during plastic deformation, with ~50% more grain rotation toward the tensile axis at 600 K than at 300 K.

As a whole, we hope that our modest contribution of the GTA analysis tool can have an impact by encouraging dialogue and data-sharing between the computational and experimental materials characterization communities. This analysis code will be provided to any interested researchers who would like to quantify microstructure in atomistic data files. It is our hope that any improvements will in turn be made available to the ICME community.

**ACKNOWLEDGEMENTS**

We gratefully acknowledge support from the National Science Foundation through a CAREER Award No. DMR-1255305.

**REFERENCES**

[1] J. Allison, D. Backman, L. Christodoulou: *JOM*, 2006, vol. 58, pp. 25-27.
[2] J. Allison: *JOM*, 2011, vol. 63, pp. 15-18.
[3] National Research Council, Integrated Computational Materials Engineering: A Transformational Discipline for Improved Competitiveness and National Security National Academies Press, Washington, D.C. , 2008
[4] J.H. Panchal, S.R. Kalidindi, D.L. McDowell: *Comput.-Aided Des.*, 2013, vol. 45, pp. 4-25.
[5] J.E. Spowart: *Scr. Mater.*, 2006, vol. 55, pp. 5-10.
[6] M.D. Uchic, M.A. Groeber, D.M. Dimiduk, J.P. Simmons: *Scr. Mater.*, 2006, vol. 55, pp. 23-28.
[7] M.P. Echlin, A. Mottura, C.J. Torbet, T.M. Pollock: *Rev. Sci. Instrum.*, 2012, vol. 83, pp. 023701.
[8] S. Ma, J.P. McDonald, B. Tryon, S.M. Yalisove, T.M. Pollock: *Metall. Mater. Trans. A*, 2007, vol. 38A, pp. 2349-2357.
[9] P.A. Midgley, M. Weyland: *Ultramicroscopy*, 2003, vol. 96, pp. 413-431.
[10] T.F. Kelly, M.K. Miller: *Rev. Sci. Instrum.*, 2007, vol. 78, pp. 031101.
[11] H.C. Huang, H. Van Swygenhoven: *MRS Bull.*, 2009, vol. 34, pp. 160-162.
[12] H.L. Wang, L.A. Zepeda-Ruiz, G.H. Gilmer, M. Upmanyu: *Nat. Commun.*, 2013, vol. 4, pp. 1956.
[13] B.Q. Cheng, A.H.W. Ngan: *Comput. Mater. Sci.*, 2013, vol. 74, pp. 1-11.
[14] B. Sadigh, P. Erhart, A. Stukowski, A. Caro, E. Martinez, L. Zepeda-Ruiz: *Phys. Rev. B*, 2012, vol. 85, pp. 184203.
[15] C.-W. Pao, S.M. Foiles, E.B. Webb, III, D.J. Srolovitz, J.A. Floro: *Phys. Rev. B*, 2009, vol. 79, pp. 224113.
[16] J. Schiotz: *Mater. Sci. Eng. A*, 2004, vol. 375, pp. 975-979.
[17] J. Monk, D. Farkas: *Phys. Rev. B*, 2007, vol. 75, pp. 045414.
[18] L. Li, J.L. Shao, Y.F. Li, S.Q. Duan, J.Q. Liang: *Chin. Phys. B*, 2012, vol. 21, pp. 026402.
[19] A.C. Lund, C.A. Schuh: *Appl. Phys. Lett.*, 2003, vol. 82, pp. 2017-2019.
[20] A. Stukowski, K. Albe: *Modell. Simul. Mater. Sci. Eng.*, 2010, vol. 18, pp. 085001.
[21] A. Stukowski, V.V. Bulatov, A. Arsenlis: *Modell. Simul. Mater. Sci. Eng.*, 2012, vol. 20, pp. 085007.
[22] T. Xu, M. Li: *Philos. Mag.*, 2010, vol. 90, pp. 2191-2222.
[23] C.D. Barrett, M.A. Tschopp, H. El Kadiri: *Scr. Mater.*, 2012, vol. 66, pp. 666-669.
[24] G.J. Tucker, J.A. Zimmerman, D.L. McDowell: *Int. J. Eng. Sci.*, 2011, vol. 49, pp. 1424-1434.
[25] P.M. Derlet, S. Van Petegem, H. Van Swygenhoven: *Phys. Rev. B.*, 2005, vol. 71, pp. 024114.
[26] S.P. Coleman, D.E. Spearot, L. Capolungo: *Model. Simul. Mater. Sci. Eng.*, 2013, vol. 21, pp. 055020.
[27] G.J. Tucker, S.M. Foiles: *Mater. Sci. Eng. A*, 2013, vol. 571, pp. 207-214.
[28] C.L. Kelchner, S.J. Plimpton, J.C. Hamilton: *Phys. Rev. B*, 1998, vol. 58, pp. 11085-11088.
[29] D. Faken, H. Jonsson: *Comput. Mater. Sci.*, 1994, vol. 2, pp. 279-286.
[30] H. Tsuzuki, P.S. Branicio, J.P. Rino: *Comput. Phys. Commun.*, 2007, vol. 177, pp. 518-523.
[31] G.J. Ackland, A.P. Jones: *Phys. Rev. B*, 2006, vol. 73, pp. 054104.
[32] A. Stukowski: *Modell. Simul. Mater. Sci. Eng.*, 2012, vol. 20, pp. 045021.
[33] S. Plimpton: *J. Comput. Phys.*, 1995, vol. 117, pp. 1-19.
[34] J.J. Gilvarry: *Phys. Rev.*, 1956, vol. 102, pp. 308-316.
[35] Y. Mishin, D. Farkas, M.J. Mehl, D.A. Papaconstantopoulos: *Phys. Rev. B*, 1999, vol. 59, pp. 3393-3407.
[36] A. Stukowski: *Modell. Simul. Mater. Sci. Eng.*, 2010, vol. 18, pp. 015012.
[37] D.E. Spearot, K.I. Jacob, D.L. McDowell: *Acta Mater.*, 2005, vol. 53, pp. 3579-3589.
[38] J.W. Cahn, Y. Mishin, A. Suzuki: *Acta Mater.*, 2006, vol. 54, pp. 4953-4975.
[39] K.S. Kumar, H. Van Swygenhoven, S. Suresh: *Acta Mater.*, 2003, vol. 51, pp. 5743-5774.

[40] D. Farkas, L. Patrick: *Philos. Mag.*, 2009, vol. 89, pp. 3435-3450.
[41] V. Yamakov, D. Wolf, S.R. Phillpot, A.K. Mukherjee, H. Gleiter: *Nat. Mater.*, 2002, vol. 1, pp. 45-48.
[42] A.C. Lund, C.A. Schuh: *Acta Mater.*, 2005, vol. 53, pp. 3193-3205.
[43] E. Bitzek, P.M. Derlet, P.M. Anderson, H. Van Swygenhoven: *Acta Mater.*, 2008, vol. 56, pp. 4846-4857.
[44] T.J. Rupert, C.A. Schuh: *Philos. Mag. Lett.*, 2012, vol. 92, pp. 20-28.
[45] E.D. Tabachnikova, A.V. Podolskiy, V.Z. Bengus, S.N. Smirnov, M.I. Bidylo, H. Li, P.K. Liaw, H. Choo, K. Csach, J. Miskuf: *Mater. Sci. Eng. A*, 2009, vol. 503, pp. 110-113.
[46] J. Schiotz, T. Vegge, F.D. Di Tolla, K.W. Jacobsen: *Phys. Rev. B*, 1999, vol. 60, pp. 11971-11983.
[47] H. VanSwygenhoven, A. Caro: *Nanostruct. Mater.*, 1997, vol. 9, pp. 669-672.

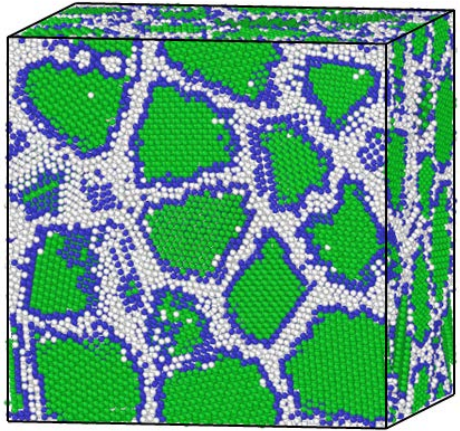

**Figure 1.** Nanocrystalline Al atomistic sample, with atoms separated into grain interior (green), grain edge (blue), and non-crystalline (white). CSP was used to find defective atoms, and the non-crystalline atoms are all grain boundary atoms since there is no stored dislocation debris. (A color version of this figure is available online.)

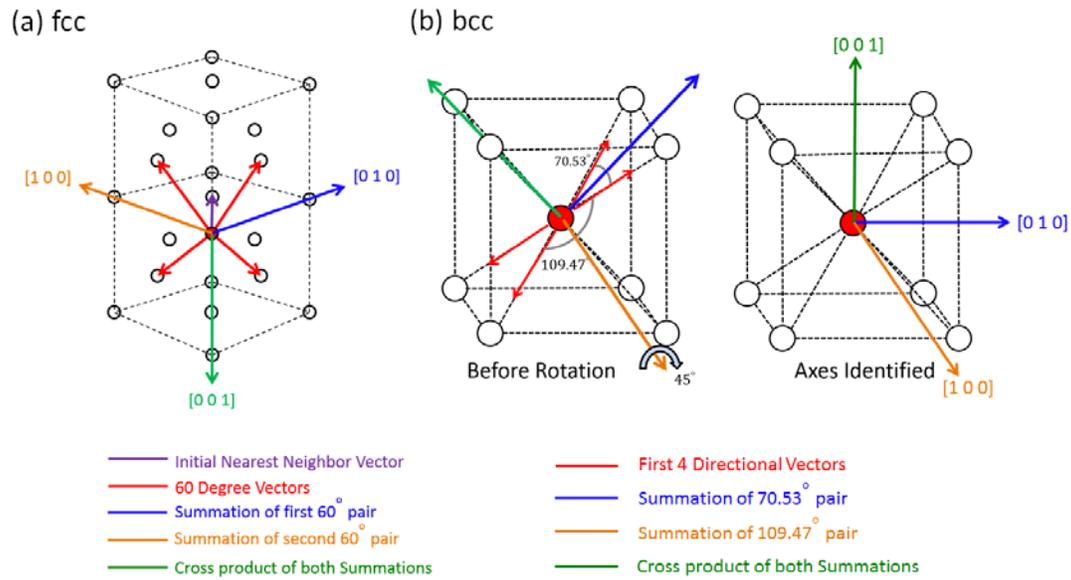

**Figure 2.** (a) A schematic of atoms in two stacked fcc unit cells which illustrates the process of calculating the local crystallographic orientation of an atom. The inverse of the three calculated <100> directions must also be taken to find all six <100> directions. (b) A similar schematic illustration of the process used to find the orientation of atoms in bcc environments. Again, all <100> directions are found and uses to store orientation information. (A color version of this figure is available online.)

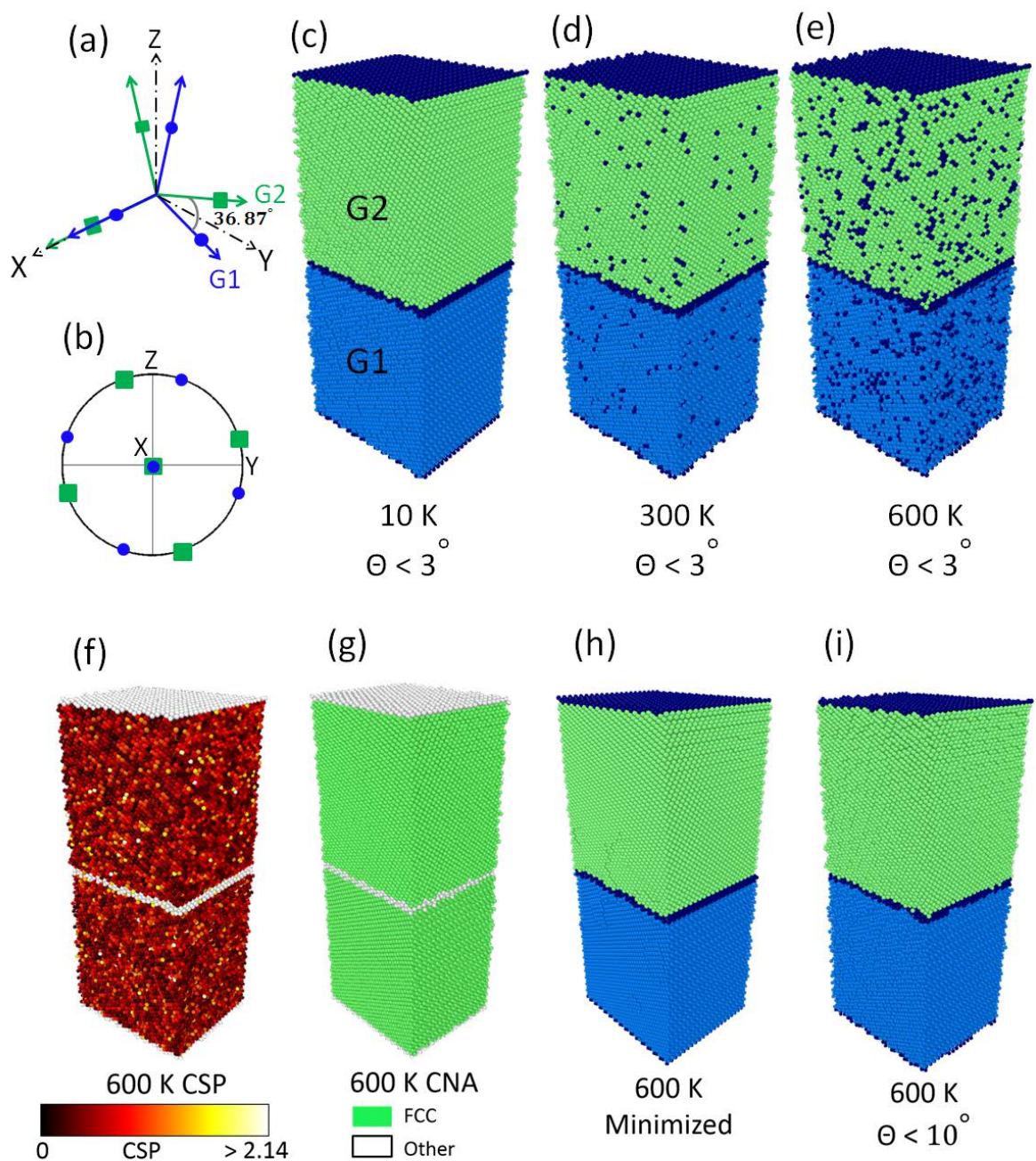

**Figure 3.** (a) and (b) show the <100> axes of the two grains from a Ni bicrystal in a vector schematic and a pole figure, respectively. (c)-(e) Samples colored according to grain number, using an orientation-cutoff angle (θ) of 3°, with light blue for G1 and green for G2, show increasing numbers of dark blue, unindexed atoms as temperature is increased. (f) and (g) show that CSP and CNA are not always good indicators of those atoms which will have large variations in local orientation. A conjugate gradient minimization (h) or an increase in the orientation-cutoff angle (i) will reduce the number of unindexed atoms. (A color version of this figure is available online.)

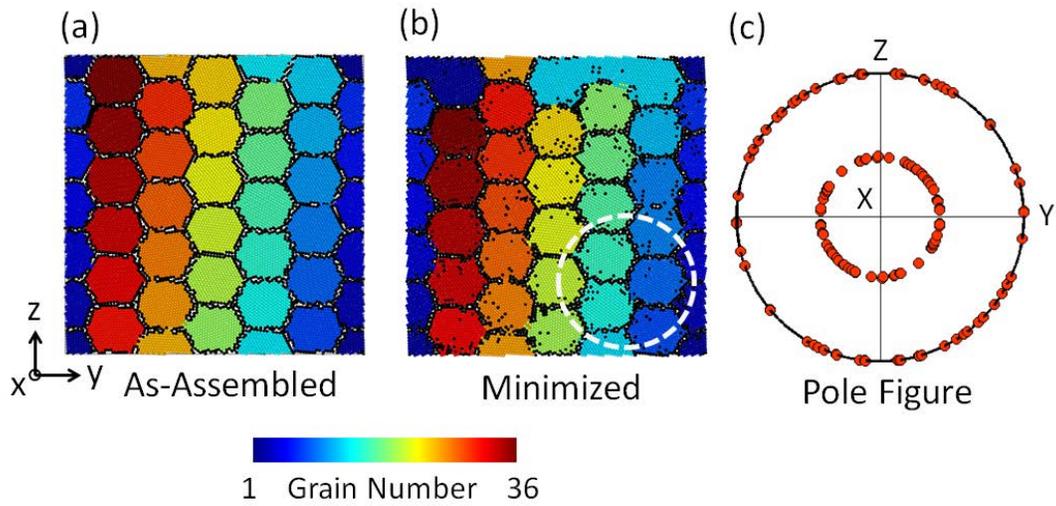

**Figure 4.** A columnar grained Al sample consisting of 36 grains, all with their {110} crystal planes oriented in the X direction. Atoms are labeled by grain number for (a) the as-assembled sample and (b) after energy minimization. (c) A {100} pole figure along the X-axis of the simulation cell reveals the sample texture.

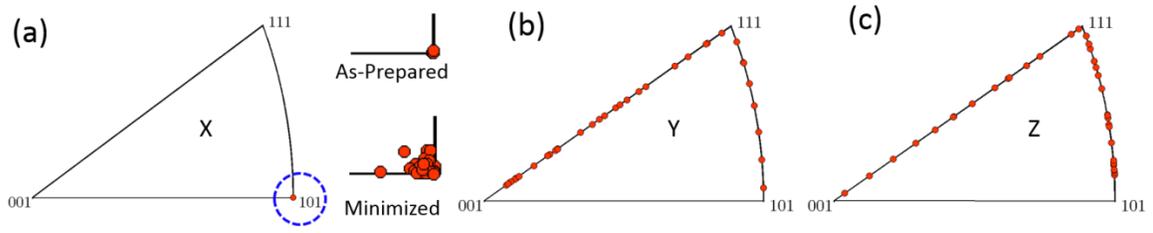

**Figure 5.** Inverse pole figures taken from the columnar grained Al sample, with each point on the triangle corresponding to a different grain. Along the X-axis of the sample, all grains have a {100} texture, while the other directions show a distribution of orientations. Energy minimization changes the out-of-plane orientation by a maximum of 0.1° and much less for most grains.

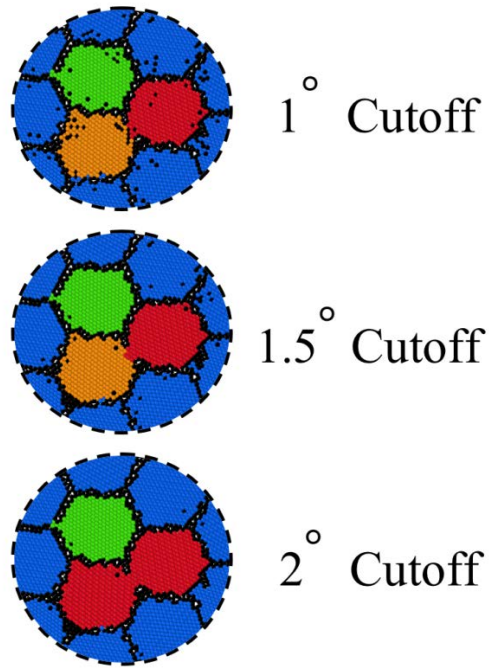

**Figure 6. A collection of three grains within the columnar grained sample. As orientation-cutoff angle is increased, the number of unindexed atoms (black atoms) is reduced significantly. However, increasing the orientation-cutoff can also lead to two grains being identified as one, as shown in the case of a 2° cutoff angle.**

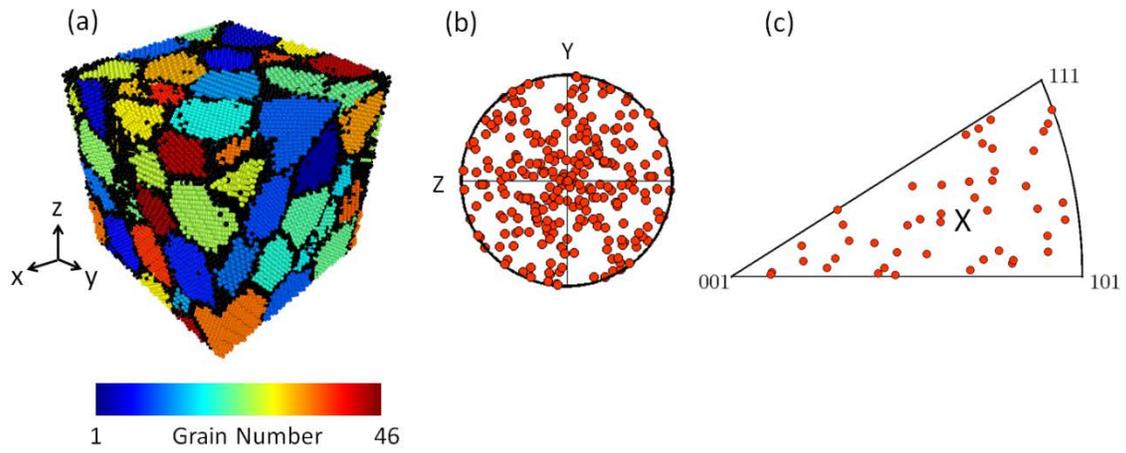

**Figure 7.** (a) A polycrystalline Al sample, with 46 randomly oriented grains. Atoms are colored according to their grain number. The random orientation is expressed in both (b) a {100} pole figure and (c) an inverse pole figure along the X-axis of the sample coordinates.

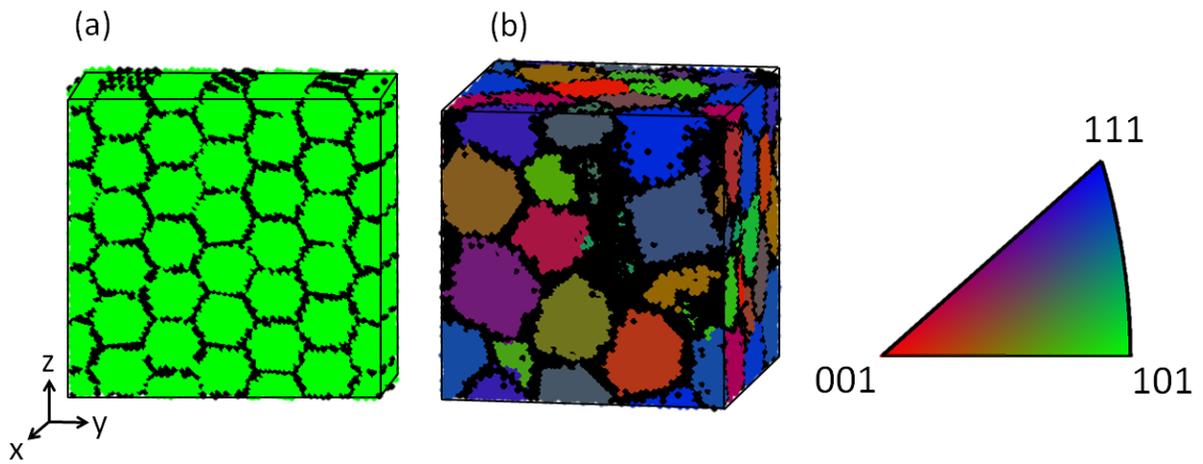

**Figure 8.** (a) Orientation map from the X-axis of the columnar sample, showing the expected {110} texture. (b) Orientation map from the X-axis of the random polycrystalline sample, showing the expected random texture.

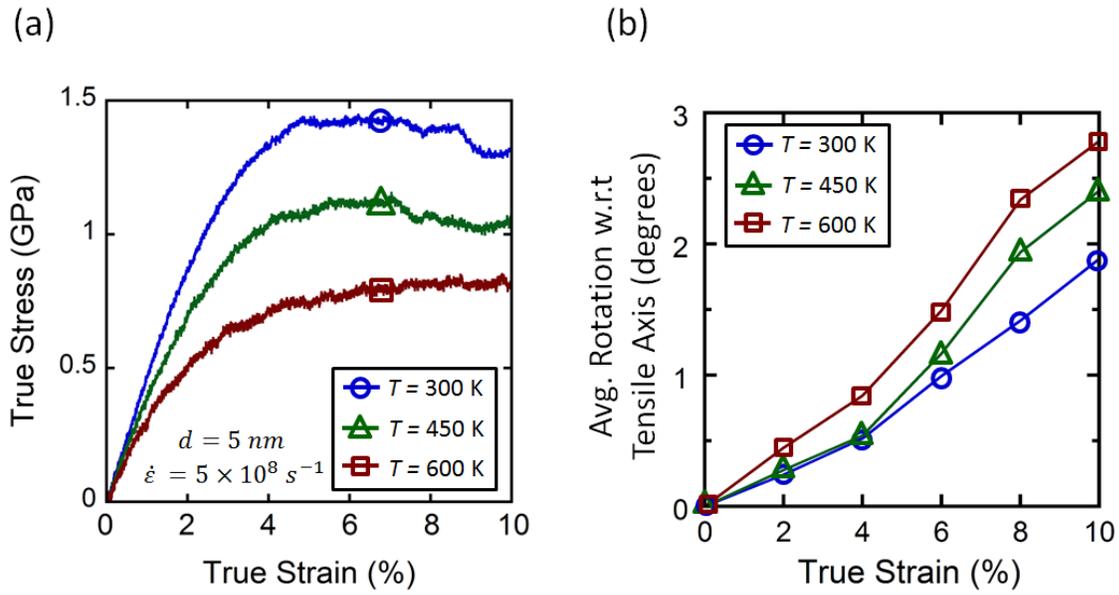

**Figure 9. (a) Tensile stress-strain curves for nanocrystalline Al samples with a mean grain size of 5 nm, tested at different temperatures. (b) Average grain rotation from starting configuration, measured as the angle with respect to the tensile axis. Increasing temperature from 300 K to 600 K leads to a ~50% increase in average grain rotation.**

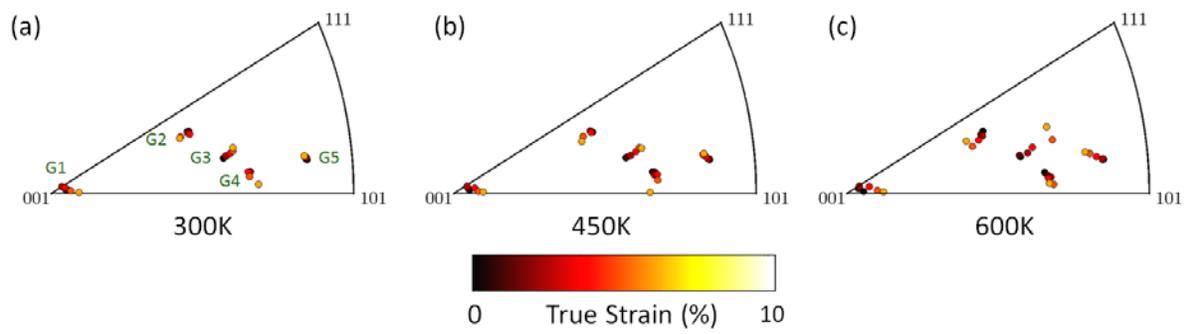

**Figure 10. Inverse pole figures for a nanocrystalline Al sample deformed at three different temperatures, showing 5 different grains and tracking their orientation evolution as a function of time. While the direction of rotation stays the same, the amount of rotation increases with increasing temperature.**

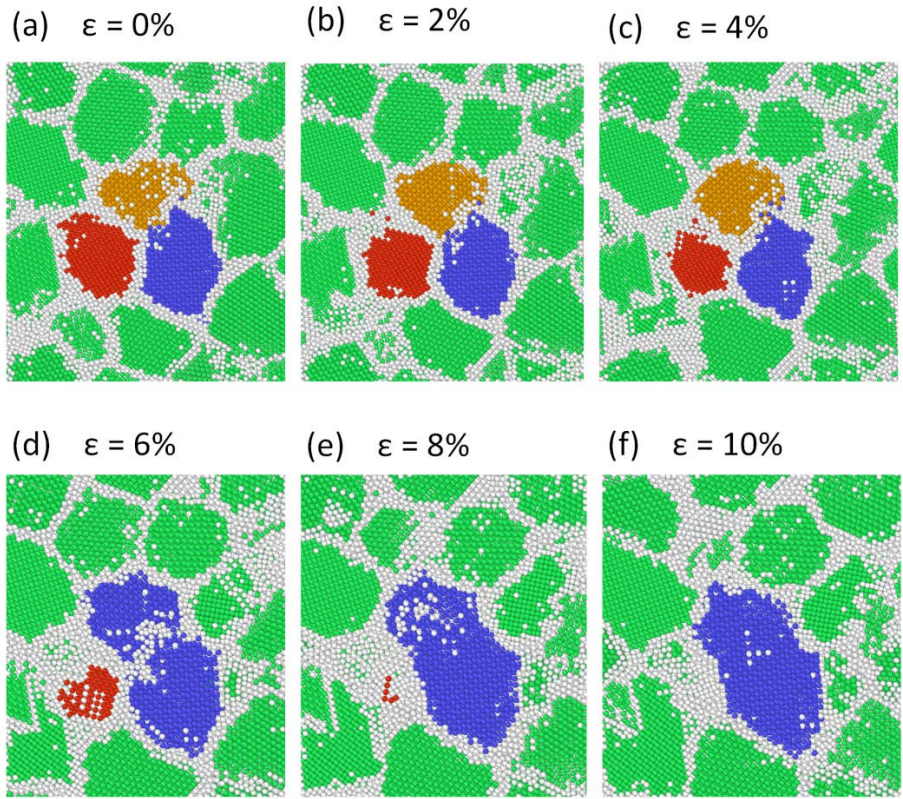

**Figure 11. Tracking of grain coalescence during tensile loading of nanocrystalline Al. Three grains are identified in (a). As strain is applied, the gold and blue grains rotate toward each other and merge, while the red grain slides into the page.**